\documentclass[sigconf,nonacm]{acmart}
\renewcommand\footnotetextcopyrightpermission[1]{} 
\usepackage{graphicx}
\usepackage{tabularx}
\usepackage{amsmath}
\usepackage{url}
\usepackage{multirow}
\usepackage{array}
\usepackage{moreverb}
\usepackage{rotating}
\usepackage{tcolorbox}
\usepackage{dialogue}
\usepackage{float}
\usepackage{cleveref}
\usepackage{footnote}
\usepackage{caption}
\usepackage{caption}
\usepackage{subcaption}
\usepackage[htt]{hyphenat}
\usepackage{dblfloatfix}

\newfloat{excerpt}{h}{lob}
\floatname{excerpt}{Excerpt}
\Crefname{excerpt}{Excerpt}{Excerpts}
\newcolumntype{Y}{>{\centering\arraybackslash}X}

\begin{document}

\title{The Apiza Corpus:\\ API Usage Dialogues with a Simulated Virtual Assistant}

\author{Zachary Eberhart}
\email{zeberhar@nd.edu}
\affiliation{%
  \institution{University of Notre Dame}
  \streetaddress{Dept. of Computer Science and Engineering}
  \city{Notre Dame}
  \state{Indiana}
  \postcode{46656}
}

\author{Aakash Bansal}
\email{abansal1@nd.edu}
\affiliation{%
  \institution{University of Notre Dame}
  \streetaddress{Dept. of Computer Science and Engineering}
  \city{Notre Dame}
  \state{Indiana}
  \postcode{46656}
}

\author{Collin McMillan}
\email{cmc@nd.edu}
\affiliation{%
  \institution{University of Notre Dame}
  \streetaddress{Dept. of Computer Science and Engineering}
  \city{Notre Dame}
  \state{Indiana}
  \postcode{46656}
}


\begin{abstract}
Virtual assistant technology has the potential to make a significant impact in the field of software engineering. However, few SE-related datasets exist that would be suitable for the design or training of a virtual assistant. To help lay the groundwork for a hypothetical virtual assistant for API usage, we designed and conducted a Wizard-of-Oz study to gather this crucial data. We hired 30 professional programmers to complete a series of programming tasks by interacting with a simulated virtual assistant. Unbeknownst to the programmers, the virtual assistant was actually operated by another human expert. In this report, we describe our experimental methodology and summarize the results of the study.
\end{abstract}

\maketitle
\thispagestyle{plain}

\vspace{-.2cm}
\section{Introduction}
\label{sec:intro}

Virtual Assistants (VAs) are software systems that interact with human users via natural language and perform tasks at the request of those users.  VAs for everyday tasks (e.g., Cortana, Alexa, Siri) are proliferating after a period of heavy investment -- a confluence of sufficient training data, advancements in artificial intelligence, and consumer demand have fed rapid growth~\cite{white2018skill}. At the same time, researchers are working to develop virtual assistants for more specialty applications in fields like medicine~\cite{white2018skill}, education~\cite{white2018skill}, and recently, software engineering~\cite{boehm2006view, robillard2017demand}. 

In software engineering (SE), one task that cries out for help from virtual assistants is API usage: programmers trying to use an unfamiliar API to build a new software program.  As Robillard~\emph{et al.}~\cite{robillard2017demand} point out, API usage is a high value target for VAs due to the high complexity of the task, and a tendency to need the same types of information about different APIs~\cite{robillard2009makes}.  Plus, the authors of APIs are often not available to answer questions (e.g. for free APIs found online), and web support (e.g. StackOverflow) is neither a guarantee nor immediately available, which makes the rapid help a VA can provide more valuable.

And yet, working relevant VA technology for programmers remains something of a ``holy grail.'' This is due, in part, to the lack of specialized datasets that are needed to design and train a virtual assistant. Modern AI systems ``learn'' to recognize and reproduce conversational patterns from large, annotated datasets of relevant dialogue. The language and strategies employed in dialogue varies across different domains, meaning that the training data provided to a system must be task-specific. Very few of these datasets exist in the field of software engineering. A 2015 survey~\cite{serban2015survey} found none related to SE tasks, and since that time only one has been published to our knowledge, targeting the task of bug repair~\cite{wood2018detecting}.  

A recent book by Reiser and Lemon~\cite{rieser2011reinforcement} provides clear guidance on how to efficiently collect the data needed to design a task-specific VA. In short, they explain that a highly-effective method to kick start development of VAs is to 1) conduct ``Wizard of Oz'' (WoZ) experiments to collect conversations between humans and (simulated) VAs and 2) annotate each utterance in the conversations with ``Dialogue Act'' (DA) types. In WoZ experiments, participants interact with a VA to complete a task, but they are unaware that the VA is actually operated by a human ``wizard''.  The deception is necessary because people communicate differently with machines than they do with other humans~\cite{dahlback1993wizard} and our objective is to create data for a machine to learn strategies to converse with humans.
The key element of these strategies are dialogue acts: spoken or written utterances that accomplish specific goals in a conversation. A VA learns from training data with annotated dialogue acts to recognize when a human is e.g. asking a question or providing information.

Following the procedure outlined by Reiser and Lemon, we conducted WoZ experiments designed to lay a foundation for the creation of virtual assistants for API usage.  We hired 30 professional programmers to complete programming tasks with the help of a ``virtual assistant,'' which was operated by a human wizard.  The programmers conversed with the virtual assistant, though they were not aware that it was operated by a human.  Each programming session lasted approximately 90 minutes. We then annotated the dialogue acts in all 30 conversations across four relevant dimensions. We make all data available via an online repository.

\begin{excerpt}[b!]
\begin{tcolorbox}[left=5pt,right=-15pt,top=5pt,bottom=5pt]
\begin{flushleft}
\begin{dialogue}
\speak{Pro} allegro keyboard input
\vspace{0.1cm}
\speak{Wiz} You can save the state of the keyboard specified at the time the function is called into the structure pointed to by \texttt{ret\_state}, using \texttt{al\_get\_keyboard\_state}
\vspace{0.1cm}
\speak{Pro} Whats the function signature for \texttt{al\_get\_keyboard\_state} 
\vspace{0.1cm}
\speak{Wiz} \texttt{void al\_get\_keyboard\_state( ALLEGRO\_KEYBOARD\_STATE *ret\_state)}
\end{dialogue}
\end{flushleft}
\end{tcolorbox}
\caption{A typical exchange between a programmer (``PRO'') and wizard (``WIZ'') in our WoZ API usage corpus.}
\label{exc:basic}
\vspace{-.3cm}

\end{excerpt}

\section{Wizard of Oz Experiments}

\begin{table*}[t!]
	\caption{The different dimensions along which the corpus was annotated.} 
	\vspace{-.2cm}
	\centering
	\begin{tabularx}{\textwidth}{lXX}
		\toprule
		Dialogue Act Dimension & \multicolumn{1}{c}{Summary} & \multicolumn{1}{c}{Examples} \\
		\midrule
		Illocutionary Dialogue Act Types  & Set of 14 labels describing the forward-facing illocutionary force of an utterance.& \texttt{INFORM, ELICIT-INFORM, SUGGEST, ...}  \\
		\vspace{0.1cm}
		API Dialogue Act Types     & Set of 11 labels describing the API information referenced in an utterance.& \texttt{FUNCTIONALITY, PATTERNS, EXAMPLES, CONTROL FLOW, ...}  \\
		\vspace{0.1cm}
		Backward-Facing Dialogue Act Types     & Set of 6 labels describing the relationship of an utterance to a previous utterance.    & \texttt{POSITIVE, NEGATIVE, PARTIAL, ...}  \\
		\vspace{0.1cm}
		Traceability   & Set of all relevant components in an API.    & \texttt{ssh\_session, ALLEGRO\_KEY\_DOWN, ssh\_disconnect(ssh\_session session)...}  \\
		\bottomrule
	\end{tabularx}
	\label{tab:DAtypes}
	\vspace{-.2cm}
\end{table*}

A ``Wizard of Oz'' (WoZ) experiment is one in which a human (the user) interacts with a computer interface that the human believes is automated, but is in fact operated by another person (the wizard)~\cite{dahlback1993wizard}.  The purpose of a WoZ experiment is to collect conversation data unbiased by the niceties of human interaction: people interact with machines differently than they do with other people~\cite{schatzmann2007agenda}.  These unbiased conversation data are invaluable for kickstarting the process of building an interactive dialogue system.

We designed two scenarios in which programmers were asked to complete programming tasks using an API that was unfamiliar to them. The first scenario used the LibSSH networking library, while the second used the Allegro multimedia library. In lieu of documentation, we introduced the programmers to an ``experimental virtual assistant,'' which we named \textit{Apiza}. Unbeknownst to the programmers, Apiza was controlled by a human ``wizard.''

\vspace{-.2cm}
\subsection{Methodology}

Each study session involved two participants: a ``programmer'' and a ``wizard.'' At the start of each session, we instructed the programmer to open a virtual machine testing environment and login to a Slack channel for communication. At the same time, we had the wizard login to the same Slack channel using an account named ``Apiza''.

We provided the programmer with a full description of the scenario, consisting of specific tasks to complete with the unfamiliar API. We asked that all API-related questions be directed via Slack text messages to our ``experimental virtual assistant,'' Apiza. We described Apiza as an ``advanced AI,'' able to carry out ``organic, natural-language conversation'' and discuss ``both high-level and low-level functionality.'' We did not allow the programmer access to the API's documentation -- a constraint necessary to force programmers out of their habits and into using the experimental tool~\cite{riek2012wizard}.

Once the programmer confirmed that he or she understood the description and tasks, we started a timer and instructed the programmer to begin. For the next 90 minutes, the programmer worked through as many of the tasks as he or she could. Throughout, the programmer sent messages to the wizard, who answered them as quickly and correctly as he or she could. We instructed the wizard that his or her responses didn't need to seem ``robotic,'' but at no point was the wizard to reveal that Apiza was a human.  

When the time ran out or the programmer finished all of the tasks, we instructed the programmer to stop working. We then asked the programmer to send us his or her code and internet searches and to complete an exit survey on user satisfaction.

\vspace{-.1cm}
\subsection{Results}

\Cref{exc:basic} demonstrates a typical interaction that occurred in the WoZ API usage dialogues. We collected 30 dialogues in total. Each ``programmer'' was involved in only one dialogue, while each ``wizard'' participated in 1-10 dialogues. Across all dialogues, participants collectively generated 1927 Slack messages Wizards and programmers sent similar quantities of messages, averaging to 31.8 messages/dialogue sent by programmers and 33.1 messages/dialogue sent by wizards. The dialogues contain a total of 47928 word tokens with a vocabulary size of 3190 words. Wizards used considerably more words (41185) and drew from a larger vocabulary (2988) than programmers, who used 6743 total words and 880 unique words.

We made four key observations about the dialogues. First, we found that the programmers had generally favorable opinions of Apiza, as evidenced by their responses to the user satisfaction survey. They were particularly impressed by the system's ability to comprehend their messages. Programmers indicated that Apiza's biggest weakness was its slow response time. Second, we observed that not every programmer attempted or completed every task. We speculate that task attempt and success rates are related to both the technical ability of the programmers, and the content and quality of the interactions with the wizards. Third, we observed that programmers demonstrated habits that they may have learned by interacting with other VAs and tools. For instance, programmers frequently avoided the use of pronouns to refer to recently mentioned API components, opting instead to repeat full function or variable names. Finally, we observed that speakers often expressed multiple, separate ideas and intentions within a single message. Identifying these separate utterances and parsing their meanings are key challenges a virtual assistant needs to overcome. 

\vspace{-.1cm}

\section{Dialogue Act Annotation}

\begin{figure*}[t]
  \includegraphics[width=\linewidth]{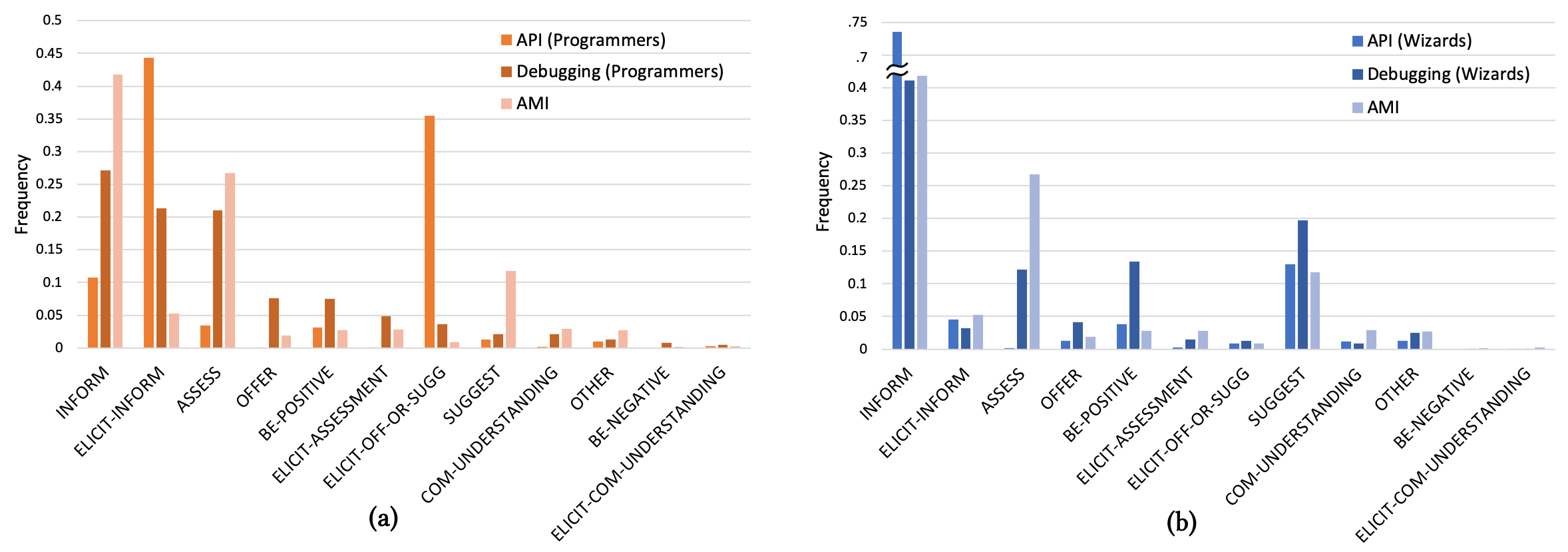}
  \vspace{-.7cm}
   \caption{Frequencies of illocutionary dialogue act types in our WoZ API usage corpus, the debugging corpus by Wood~\emph{et al.}~\cite{wood2018detecting}, and the AMI meeting corpus~\cite{mccowan2005ami}. Subfigure (a) shows programmer frequencies, while subfigure (b) shows wizard frequencies.}
  \label{fig:DAdist}
  \vspace{-.2cm}
\end{figure*} 
\vspace{-0.1cm}

\begin{figure*}[t]
\centering
\begin{minipage}[t]{\columnwidth}
	\centering
  \includegraphics[width=\linewidth]{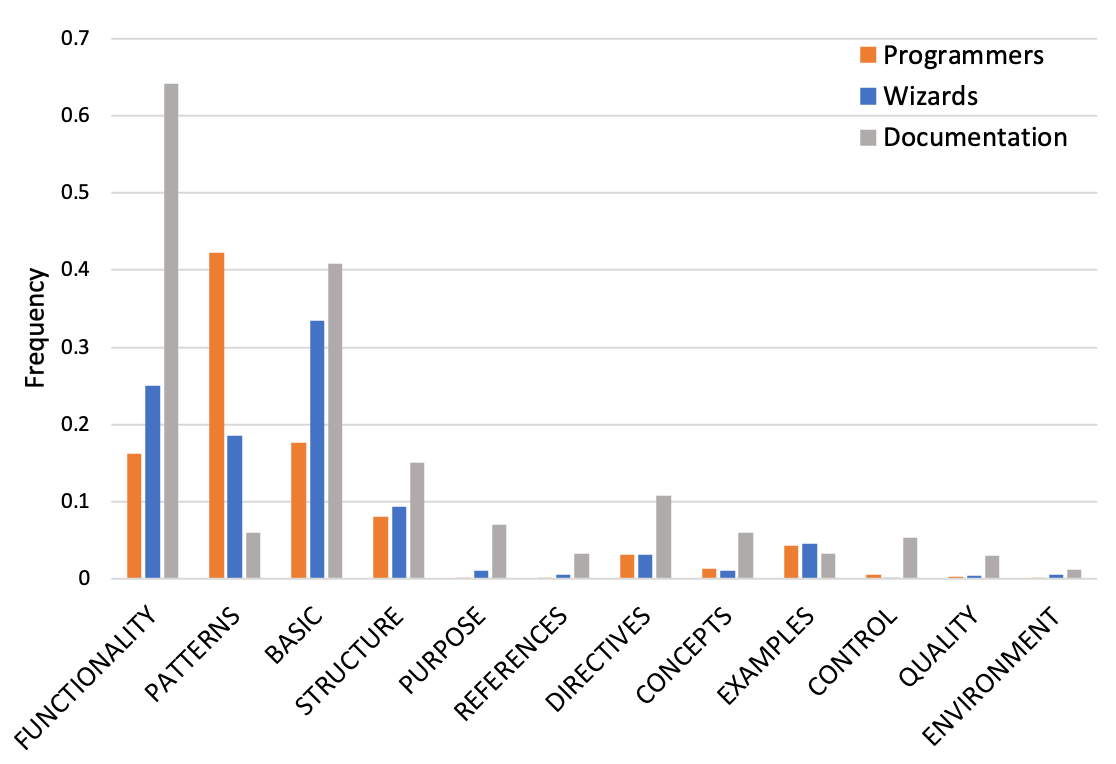}
  \vspace{-.6cm}

   \caption{Frequencies of API dialogue act types in our WoZ API usage corpus and the documentation for JDK 6~\cite{maalej2013patterns}}

  \label{fig:APIdist}
\end{minipage}
\hfill
\begin{minipage}[t]{\columnwidth}
	\centering
  \includegraphics[width=\linewidth]{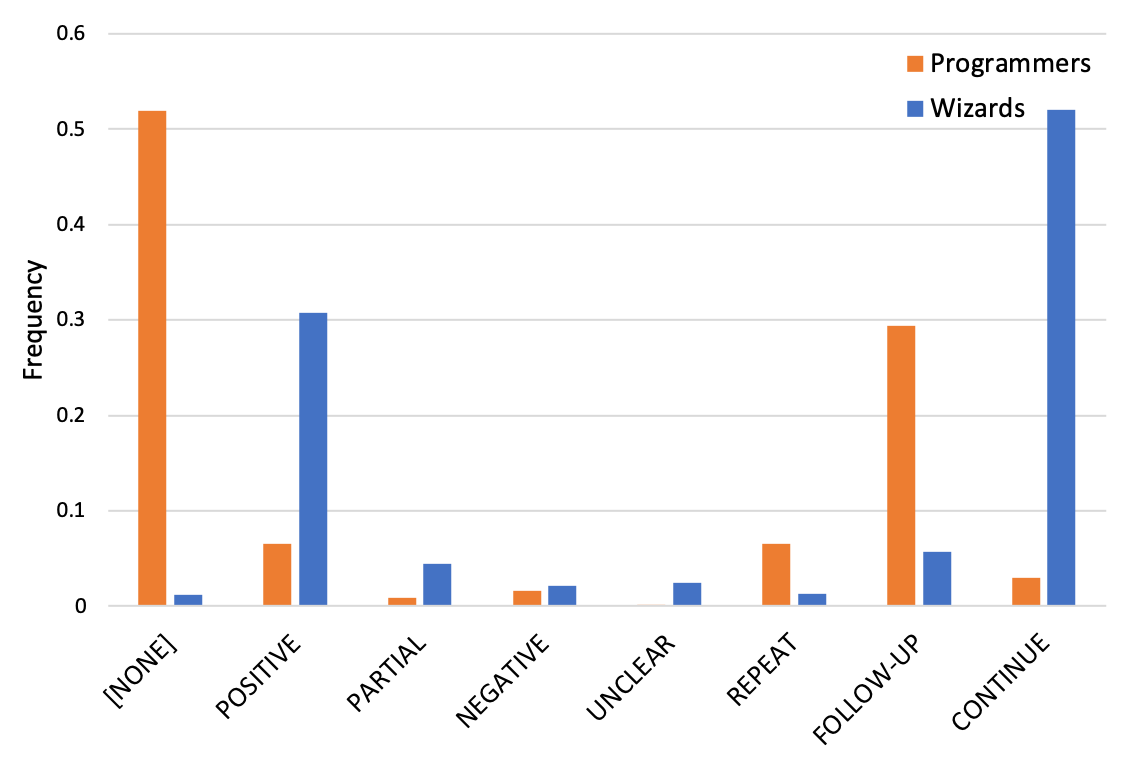}
  \vspace{-.6cm}
   \caption{Frequencies of backward-facing dialogue act types in our WoZ API usage corpus.}
  \label{fig:BFdiff}
\end{minipage}
\vspace{-.2cm}
\end{figure*}

We annotated the dialogue acts in the WoZ corpus along the four dimensions shown in \Cref{tab:DAtypes}: \emph{illocutionary} DA types, \emph{API} DA types, \emph{backward-facing} DA types, and \emph{traceability}. Illocutionary DA types express the illocutionary intent behind utterances, revealing the conversational ``flow" of the API dialogues. API DA types describe what types of API knowledge (such as functionality, usage patterns, or examples) are addressed in each utterance. Backward-facing DA types capture the more complex structural relationships between dialogue acts (e.g. when a programmer follows up on a questions, or a wizard asks a clarifying question). Traceability identifies and tracks the specific API components that are addressed throughout the dialogues. We chose these dimensions to address the key tasks an API usage VA needs to perform: identifying the content, context, and meaning of a user's utterance and responding appropriately.

\vspace{-.1cm}

\subsection{Methodology}

We annotated each dimension according to a particular annotation scheme. We annotated illocutionary DA types using the scheme from the AMI corpus~\cite{mccowan2005ami}, a large, annotated corpus of simulated business meetings. This scheme was also used by Wood~\emph{et al.}~\cite{wood2018detecting} to label their WoZ debugging dataset. We annotated API DA types using the ``taxonomy of API knowledge types,'' a scheme proposed by Mallej and Robillard~\cite{maalej2013patterns} that was used to categorize documentation in large APIs, such as JDK 6. We annotated backward-facing DA types using another layer of the AMI corpus~\cite{mccowan2005ami} annotation scheme. We introduced a small number of classes to the core label set to provide for additional granularity in backward-facing structural relationships. Finally, we annotated traceability using the names of specific API components that are referenced in the dialogues. 
For the first three annotation schemes, we followed detailed guides produced by the authors of the annotation schemes to determine the correct labels and resolve any ambiguities.  

\vspace{-.2cm}

\subsection{Results}

\Cref{fig:DAdist} shows the distribution of illocutionary DA types in our corpus, compared to that in the WoZ debugging corpus~\cite{wood2018detecting} and the AMI meeting corpus~\cite{mccowan2005ami}. We found that compared to the other corpora, programmers elicited more information and suggestions and wizards provided more information and suggestions.

\Cref{fig:APIdist} shows the distribution of API DA types in our corpus, compared to those in the JDK 6 documentation~\cite{maalej2013patterns}. Programmers primarily asked questions relating to Pattern, Basic, and Functionality information. Compared to the JDK 6 documentation, there is substantially more focus on Pattern information. 

\Cref{fig:BFdiff} shows the distribution of backward-facing DA types in our corpus. Programmers frequently asked follow-up questions, while wizards primarily provided direct (``positive'') answers. 

Finally, we identified 157 specific API components referenced across 2430 utterances. The wizards referred to specific API components more often than the programmers, and a slightly broader range of API components were identified in the Allegro scenario than in the LibSSH scenario.

\vspace{-.2cm}
\section{Data Repository}
To facilitate reproducibility and future research, we have made all data related to the experimental design, experimental results, and
dialogue act annotations available via an online repository:
\\
\vspace{-.2cm}
\\
\textbf{https://github.com/ApizaCorpus/ApizaCorpus}
\vspace{-.1cm}



\bibliographystyle{ACM-Reference-Format}
\bibliography{biblio}

\end{document}